\documentstyle[12pt]{article}
 \begin{document} 


\newskip\humongous \humongous=0pt plus 1000pt minus 1000pt
\def\caja{\mathsurround=0pt}
\def\eqalign#1{\,\vcenter{\openup1\jot \caja
 \ialign{\strut \hfil$\displaystyle{##}$&$
 \displaystyle{{}##}$\hfil\crcr#1\crcr}}\,}
\newif\ifdtup
\def\panorama{\global\dtuptrue \openup1\jot \caja
 \everycr{\noalign{\ifdtup \global\dtupfalse
 \vskip-\lineskiplimit \vskip\normallineskiplimit
 \else \penalty\interdisplaylinepenalty \fi}}}
\def\eqalignno#1{\panorama \tabskip=\humongous
 \halign to\displaywidth{\hfil$\displaystyle{##}$
 \tabskip=0pt&$\displaystyle{{}##}$\hfil
 \tabskip=\humongous&\llap{$##$}\tabskip=0pt
 \crcr#1\crcr}}
\jot = 1.5ex
\def\baselinestretch{1.2}
\parskip 5pt plus 1pt
\catcode`\@=11
\@addtoreset{equation}{section}
\def\theequation{\arabic{section}.\arabic{equation}}
\def\@normalsize{\@setsize\normalsize{15pt}\xiipt\@xiipt
\abovedisplayskip 14pt plus3pt minus3pt%
\belowdisplayskip \abovedisplayskip
\abovedisplayshortskip \z@ plus3pt%
\belowdisplayshortskip 7pt plus3.5pt minus0pt}
\def\small{\@setsize\small{13.6pt}\xipt\@xipt
\abovedisplayskip 13pt plus3pt minus3pt%
\belowdisplayskip \abovedisplayskip
\abovedisplayshortskip \z@ plus3pt%
\belowdisplayshortskip 7pt plus3.5pt minus0pt
\def\@listi{\parsep 4.5pt plus 2pt minus 1pt
     \itemsep \parsep
     \topsep 9pt plus 3pt minus 3pt}}
\relax
\catcode`@=12
\evensidemargin 0.0in
\oddsidemargin 0.0in
\textwidth 6.0in
\textheight 8.5in
\hoffset .7 cm
\voffset -1 cm
\headsep .25in
\catcode`\@=11
\def\section{\@startsection{section}{1}{\z@}{3.5ex plus 1ex minus
   .2ex}{2.3ex plus .2ex}{\large\bf}}

\def\thesection{\arabic{section}}
\def\thesubsection{\arabic{section}.\arabic{subsection}}
\def\thesubsubsection{\arabic{subsubsection}.}
\def\appendix{\setcounter{section}{0}
 \def\thesection{Appendix \Alph{section}}
 \def\theequation{\Alph{section}.\arabic{equation}}}
\newcommand{\beq}{\begin{equation}}
\newcommand{\eeq}{\end{equation}}
\newcommand{\bea}{\begin{eqnarray}}
\newcommand{\eea}{\end{eqnarray}}
\newcommand{\beas}{\begin{eqnarray*}}
\newcommand{\eeas}{\end{eqnarray*}}
\newcommand{\defi}{\stackrel{\rm def}{=}}
\newcommand{\non}{\nonumber}
\def\de{\partial}
\def\si{\sigma}
\def\dim{\hbox{\rm dim}}
\def\sup{\hbox{\rm sup}}
\def\inf{\hbox{\rm inf}}
\def\Arg{\hbox{\rm Arg}}
\def\Im{\hbox{\rm Im}}
\def\Re{\hbox{\rm Re}}
\def\Res{\hbox{\rm Res}}
\def\Max{\hbox{\rm Max}}
\def\Abs{\hbox{\rm Abs}}
\def\infi{\infty}
\def\nrm{\parallel}

\def\1{{\rm 1}}
\def\s{{\sigma}}
\def\e{{\cal E}}
\def\o{{\cal O}}
\def\om{{{\cal O}_m}}
\def\dm{{{\partial}_m}}
\def\fp{{\rm Fp}_\mu}
\def\au{{a_1}}
\def\ad{{a_2}}

\def\gp{g^{+}}
\def\gm{g^{-}}
\def\half{{1\over 2}}
\def\res{{\rm Res}}
\begin{titlepage}
\begin{center}
{\Large
On the Short Distance Behavior of the Critical Ising Model
Perturbed by a Magnetic Field}
\end{center}
\vspace{1ex}

\begin{center}
{\large
Riccardo Guida$^{1}$ and Nicodemo Magnoli$^{2,3}$}
\end{center}
\vspace{1ex}
\begin{center}
{\it $^{1}$ CEA-Saclay, Service de Physique Th\'eorique\\
     F-91191 Gif-sur-Yvette Cedex, France}\\
{\it $^{2}$ Dipartimento di Fisica -- Universit\`a di Genova\\
     Via Dodecaneso, 33 -- 16146 Genova, Italy}\\ 
{\it $^{3}$ Istituto Nazionale di Fisica Nucleare- Sez. Genova\\
     Via Dodecaneso, 33 -- 16146 Genova, Italy}\\
\end{center}

\begin{center}
e-mail: guida@amoco.saclay.cea.fr, magnoli@ge.infn.it
\end{center}
\medskip
{\bf ABSTRACT:}
We apply here a recently developed 
approach  to
compute the  short distance corrections
to scaling  for the correlators of all primary operators of
the critical two dimensional   Ising model in a magnetic field.
The essence of the method is  the fact that  
if one deals with O.P.E. Wilson coefficients instead of
correlators, all order I.R. safe formulas can be obtained
for  the perturbative  expansion 
with respect to magnetic field.
This  approach  yields in a natural way 
the expected fractional powers of the
magnetic field, that are clearly  absent in the naive 
perturbative expression for correlators.
The  technique of the Mellin transform have been used to
compute the I.R. behavior of the regularized integrals.
As a corollary of our results, by comparing  the existing
numerical  data for  the lattice model we give 
an estimate of  the Vacuum Expectation Value of 
the energy operator, left unfixed by usual  nonperturbative approaches
(Thermodynamic Bethe Ansatz).

\vfill

\begin{flushleft}
SPhT-T96/068

GEF-Th-$7$

6-1996





\end{flushleft}
\end{titlepage}

\section{Introduction}\label{introduction}

It is well known that 
($D=2$) conformal field theories, \cite{bpz},
can be used to describe
statistical systems at the critical point, \cite{review}.
 The second step 
from this point of view is to obtain informations 
on the behavior of the system near criticality,
that is to study a conformal field theory 
perturbed by a relevant operator (i.e.  with scale
dimension $0<x<2$).

On this line a wide class of perturbations have been found to give rise to
integrable models, \cite{integrable}, i.e. models in which an infinite number of
conserved charges exists and constrains the S matrix to be factorized 
(and possibly elastic).
 Starting from the exact  knowledge of the S matrix one can investigate the 
off shell behavior of the theory  by use of the
form factor method \cite{form} that
essentially gives rise to a
 long distance expansion for correlators in terms of $e^{-M_ir}$,
$M_i$ being the masses of the 
complete theory and $r$ the argument of correlators.

Another approach to the problem \cite{CPT,dot2} is obtained 
building up in some way a 
perturbation theory around the (massless) conformal field theory, from which
one  could  get 
 informations on the short distance behavior of the complete theory
(the small adimensional
parameter being now the coupling constant times an adequate
positive  power of
$r$).
This line of research is important not only because 
it is complementary to the previous
one, but also because of its generality (integrability is not essential,
thus {\it all} perturbations can be treated if the starting point 
conformal theory is known).
However it is well known
that when the coupling is relevant,
naive perturbative expressions  for correlators 
are plagued by I.R. divergences, that constituted an unsurmountable
obstacle to the construction of a general theory of perturbations in these cases.
Nevertheless in  \cite{gm}  
some  recent results results \cite{zamo,sonoda1,mz}
(see also \cite{wilson,zinn,david} for preexisting ideas)
have been developed to give  an {\it all order I.R. safe}
general approach to describe the short distance behavior of conformal 
theories perturbed by  relevant operators.
The main idea of the method is the fact that Wilson  coefficients of
Operator Product Expansions, being short distance objects, can be taken
to have a regular, I.R. safe, perturbative expansion with respect to the
coupling.

In this paper we apply the above mentioned OPE technique  to  reconstruct 
 the short distance behavior of 
the (continuous) critical two dimensional Ising
model perturbed by a magnetic field $H$: 
clearly our results can be applied to describe
  the scaling limit of the corresponding model on the lattice.

We will derive  the first nontrivial terms
 of the expansion in the scaling 
variable $t\propto H R^{15/8}$ ($R,H$ being
respectively  the lattice spacing and lattice magnetic field) 
for 
all the correlators of the  primary operators of the
critical conformal  theory.
In particular we obtain naturally the predicted \cite{pol,zinn}
$t^{8/15}$ corrections to the spin spin correlator.

The OPE approach leaves in general  unfixed  some  constants
 that parameterize 
the Vacuum Expectation Values of the used low dimensional operators.
One of these constants has been fixed by use of known non perturbative 
results (Thermodynamic Bethe Ansatz), while the other has been
estimated by comparison with  
 the available data from numerical 
Monte Carlo simulations on the lattice.

It is worth remarking here that the
perturbative expansions
 in 
$H/|T-T_c|^{15/8}$,
(obtained starting from the massive noncritical  Ising
at zero field) that are 
 presented in     \cite{wu}, 
are clearly inapplicable when $T=T_c$: 
this is another manifestation of the I.R. divergences we spoke above.
We notice also that, being the critical Ising Model in magnetic field
 an integrable model with known S matrix, its long distance behavior can be 
computed
by use of form factor techniques \cite{delfino,delfino2}.

The plan of the paper is as follows:
in Section \ref{all} we will summarize the OPE approach, 
while in Section \ref{mellin} we describe a known mathematical technique that 
appears to be useful to regularize and disentangle I.R. divergences
of integrals, the Mellin transform.
In Section \ref{isingh} we apply  the approach to the Ising model, (all the
bulk computations are confined to \ref{computations}); in Section \ref{vev}
we use nonperturbative as well as numerical inputs  to estimate the
parameters left unfixed by the OPE method;  
found results are discussed in Section \ref{comments}.
 Conclusions are given in 
Section \ref{conclusions}.

\section{All order I.R. finite formulas}\label{all}

The goal of the method presented in \cite{gm}
(see also \cite{zamo,sonoda1,mz} and \cite{wilson,zinn,david}
 for many preexisting ideas)
 is to obtain informations on the short distance behavior of
a conformal field theory perturbed by 
 relevant operators, i.e. when  to the conformal action $S_{CFT}$
is added a perturbation of the form 
\beq \Delta S= -\int\!\! dx \sum_i {\lambda^i_B} {\o_i}_B(x), \eeq
where $\lambda^i_B$ (${\o_i}_B$) are bare couplings (operators)
and $0<x_i<2$ ($x_i\equiv x_{\o_i}$, $x_{O}$ being the scale dimension of operator $O$). 
This is achieved  by expanding in powers of the 
corresponding renormalized couplings $\lambda^i $
(Taylor expansion) the so-called
Wilson coefficients $C_{a b}^c(r, \lambda )$
 that enter in the Operator Product Expansion 
\beq\label{ope}
<\Phi_a (r) \Phi_b(0)>_\lambda \sim \sum_c 
C_{a b}^c(r, \lambda )<\Phi_c(0)>_\lambda
\eeq
for the complete theory ($\Phi_a$ 
are deformations of the conformal theory operators,
 the suffix $\lambda$ refers to the complete theory correlators;
the dependence on $\lambda^i$ of the Wilson coefficients will be omitted
in the following).

Assuming the regularity of the Wilson coefficients in 
terms  of the 
renormalized couplings $\lambda^i$ 
(minimality of the renormalization scheme), the validity of an action principle
for the derivative with respect to $\lambda^i$
 and the asymptotic convergence of OPE,
all reasonably satisfied by the complete theory, 
an I.R. finite representation has been given for 
 the $n^{th}$ derivative of $C_{ab}^c$ with respect to
 the couplings  evaluated at $\lambda^i=0$, 
involving  integrals of 
(eventually renormalized) conformal correlators.

More in detail,
by repeated differentiation with respect to couplings of 
Eq.(\ref{ope})  and by  
use of the (renormalized) action principle
\beq
\de_{\lambda^i} <[\cdots ]>_\lambda=
 \int d^2x <[{\o_i} \cdots]>_\lambda
\eeq
(square brackets $[\cdots]$ meaning renormalization)
one can first obtain some integral representation for the 
multiple derivatives of the Wilson coefficients in terms of integrated 
correlators of the complete theory.
The limit $\lambda^i\to 0 $ can then be taken safely
(after regularizing the integrals by some generic  I.R. cutoff
function $\Theta_{R}(x)$ such $\lim_{R\to \infty} \Theta_R(x)=1$)
because OPE asymptotic convergence guarantees  that  the sum of all 
(singularly I.R. divergent) contributions is actually finite.
In the limit in which all I.R.
cut off are removed ($R_n,\cdots R_1\geq R\to \infty$ below)
one also can show by dimensional arguments
 that the OPE series truncates to a simple sum, $\sum_b^*$ below.
As a result it was obtained in \cite{gm} that
  \bea & &
\sum_b^* \de_{i_1} \cdots \de_{i_n}   C_{\au \ad}^b <[\Phi_b X_\infty ]> \non\\
 &=&\lim_{R\to \infty}\{
\int \!\! dx_{1}\cdots \int \!\!
 dx_{n} \Theta_{R_{1}} (x_{1})
 \cdots \Theta_{R_{n}}(x_{n})\times  \non\\
& & 
 <[ :\o_{i_n}: \cdots :\o_{i_1}:(\Phi_\au  \Phi_\ad  
- \sum_b^* C_{\au \ad}^b \Phi_b) X_\infty]>
\non\\
&-&
\sum_b^*\de_{i_1} C_{\au \ad}^b 
\int\!\! dx_{2}\cdots \int \!\!
 dx_{n} \Theta_{R_{2}} (x_{2})
 \cdots \Theta_{R_{n}}(x_{n}) 
  \times\non\\
& &< [:\o_{i_n}: \cdots :\o_{i_2}: \Phi_b  X_\infty ]>
 + {\rm p.} +\cdots\non\\
&- & 
 \sum_b^* \de_{i_1} \cdots\de_{i_{n-1}} C_{\au \ad}^b
\int \!\!
 dx_{n} 
 \Theta_{R_{n}}(x_{n})  
 <[ :\o_{i_n}:  \Phi_b X_\infty ]> + {\rm p.} 
\}
\label{mainmainmain} \eea 
where $\de_i={\de \over \de \lambda^i}$,
${ p.}$ means all nontrivial permutations over labels of  indices
$i$,
 $X_{R'}$ is  an  arbitrary operator localized outside the sphere of radius $R'>>R\to \infty$,
 chosen to give adequate boundary conditions
(and possibly including powers of $R'$),
and $\sum_b^* $, as explained before,
 is restricted to operators $\Phi_b$  such
$x_b\le (D-x_{i_k})\cdots +(D-x_{i_n})-x_{X_R}$.

While each term in the sum  in the right  hand side 
is I.R. divergent, (and consequently requires a regularization)
 the complete right hand side   is I.R. 
finite  and independent on
 the choice of the cutoff function (what we presented above  is  an
"I.R. renormalization" more than a simple regularization).
 The structure of the 
expression is simple: the first term on the right hand side is 
a naive (generalized) perturbative one,
while the others, containing lower derivatives of Wilson coefficients,
 play the
role of nonlocal I.R. counterterms, naturally induced by the theory itself.
(A conjecture for the existence of  this mechanism 
can be found  in \cite{parisi}, in the general context of
quantum field theories. Also a rigorous proof of the conjecture
within the  MS renormalization scheme for
perturbative superrenormalizable quantum field theories
can be found in \cite{david}.) 

If the cutoff function is rotationally invariant ($\Theta_R(r)=\Theta_R(|r|)$)
only scalar operators contribute in all expressions 
above.
Also, if by dimensional analysis one knows a priori  that 
no powers $R^0$ can be obtained from the I.R. counterterms  
available in that model,
it suffices then  to compute the first (naive) term
of right hand side in some chosen I.R. regularization,
and to keep only the regular term of its asymptotic expansion when 
$R_i \to \infty$, the singularity of the naive  term
 being automatically killed by 
the I.R. counterterms\footnote{This idea was already present
in the first order computations of  \cite{zamo}.}
(which  cannot give in this case additional  finite contributions).
Roughly speaking, one expects that this picture is realized when 
the I.R. counterterms do not need   U.V. renormalization (i.e. are locally integrable),
 because in this case the renormalization scale $\mu$ is absent 
and there is no
other  possibility to obtain  an adimensional quantity from 
the  $\log R$ terms
that are expected in general  from the $R^0$  I.R. counterterms (apart from miraculous cancellations of $\log R$ divergences). 
This shortcut can be applied in particular to the case we will consider here.

We conclude by observing that the complete correlators are then
obtained by combining the derived expressions for the Wilson coefficients with 
those for the required operator Vacuum Expectation Values 
(eventually parameterized by unknown but fixed and universal
 constants, see Section \ref{vev} and discussion in
\cite{gm}).     

\subsection{Mellin transform and I.R. divergences}\label{mellin}

Having in mind that all we need are asymptotic expansions of 
(multiple) integrals in the limit of $m_i\equiv {1\over R_i}\to 0$
 above, it is worth introducing  here  a natural method to deal 
with these expansions: the Mellin transform 
\footnote{This  technique was essential for
the  analog analysis in
perturbative superrenormalizable quantum field theories
\cite{david} (and references therein).} (see \cite{wong} for a nice
introduction). 

Given a function $I(m)$ locally integrable on $(0, \infty)$,
of order $m^\alpha$ when $m\to 0$ and $m^{-\beta}$ when $m\to \infty$,
one can introduce 
 in the complex strip 
 $-\alpha <s<\beta$
 its Mellin transform
 \beq\label{mellint}
\tilde{I}(s) \equiv \int_0^\infty {dm \over m} m^s I(m).
\eeq
If $\tilde{I}(s)$ has a meromorphic 
extension in the  complex plane on the left of ${\rm Im } s= -\alpha$ 
one can derive from the inverse Mellin transform
 the asymptotic expansion,
\beq\label{asint}
I(m)\sim \sum_{i}  \res \left( m^{-s} \tilde{I}(s) \right)_{s=-\alpha_i}
\eeq
with $\alpha_1\equiv \alpha<\alpha_2<\cdots$ are the powers of $m$ in the 
asymptotic expansion of $I(m)$ (notice that multiple poles give
logarithms of $m$). In a word: singularities of Mellin transform drive
the  asymptotic expansion.

In our formulas  we have always (possibly nested) integrals of the form
\beq
I(m)=\int d^2z \Theta(m|z|) g(z)
\eeq
(we refer from here on to the rotation invariant cutoff;
notice also  the slight and obvious change of notations).
It is easy to prove the following convolution theorem:
\beq\label{mellinc}
\tilde{I}(s)= \tilde{\Theta}(s) \tilde{G}(1-s)
\eeq
where
\beq
\tilde{G}(1-s)= \int d^2z |z|^{-s} g(z)
\eeq 
is essentially the Mellin transform of $g$ (up to angular integrals).
Thus in the cases in which $\tilde{G}$ is known the integrals are
 easily done. 
Notice that if  the Mellin transform 
of $g$
strictly speaking does not exist but the 
Mellin transform of $I(m)$ is well defined,  
one can try to bypass the problem extending 
the integral ${\tilde I}(s)$ to a more general 
one, ${\tilde I}(s;\delta)$, (in which $g$ is now 
extended to some adequate $g_\delta$) 
such that it reduces
to the original for some value $\delta=\delta_0$,
 and such that convolution theorem can
 be applied: if ${\tilde I}(s;\delta)$ can be analytically continued to 
$\delta=\delta_0$, the Mellin transform of $I$
is obtained accordingly.

In particular if
 the I.R. counterterms give no finite contributions, (see Section \ref{all})
 then 
 the 
derivatives of Wilson coefficients can be unambiguously
obtained  simply by taking the 
residue of the first, "naive perturbative", term in 
(\ref{mainmainmain}).

In \ref{computations} we give the 
analytical expression of ${\tilde I}(s)$ for a general
case that allows us to compute all the 
 integrals  involved in the first order derivatives
 of 
Wilson coefficients of  primary fields of the conformal
field theory underlying the critical Ising model.

\section{Wilson Coefficients for the perturbed Ising Model}\label{isingh}
The scaling  limit of the lattice (D=2) Ising model
(see \cite{mccoy} and references therein)
 at the critical point,
$T=T_c$, $H=0$ is described by a continuous 
unitary conformal field theory,
technically ${\cal M} (3/4)$, \cite{bpz}.
The operator space of this conformal theory 
 is   generated by the primary operators
$\1,\s , \e  $
 of 
 dimension $x=0,1/8,1$.
The related fusion rules are:
\beq
[\s ][\s ] =[\1 ]+ [\e ], \qquad [\e][\e] =[\1 ], \qquad [\s ][ \e ]=[\s ];
\eeq
implying
\beq \label{fusi}
<[\e ]^k [\1 ]^m [\s]^l>=0
\eeq
for $l$ odd.
Also Kramers Wannier duality  implies
\beq\label{kw}
<[\e ]^n [\1 ]^l>=(-)^n <[\e ]^n [\1 ]^l>
.\eeq

The object of our study  will be  the extended  theory in presence of a magnetic field:
\beq
S=S_{{\rm Ising}}+\int h\s d^2z
.\eeq
Notice that  with this notation 
each derivative $\de_h$ will be realized by the insertion of the operator
$-\int d^2z \s(z)$ (action principle).

By application of the general formula (\ref{mainmainmain})
and of the selection rules (\ref{fusi}) and (\ref{kw}) we obtain
the following (nontrivial) first order order relations
\beq \label{sss}
-\de_h C_{\s \s}^{\s}(r) <\s_0 \s_{R'}>=
\int' d^2z <\s_z \s_{R'}[ \s_r\s_0 -C_{\s \s}^{\1} (r) -C_{\s \s}^{\e}(r) 
\e_0 ]>
\eeq

\beq \label{ees}
-\de_h C_{\e \e}^{\s}(r) <\s_0 \s_{R'}>=
\int' d^2z <\s_z \s_{R'}[ \e_r\e_0 -C_{\e \e}^{\1} (r) ]>
\eeq

\beq \label{se1}
-\de_h C_{\s \e}^{\1}(r) =
\int' d^2z <\s_z [ \s_r\e_0
 -C_{\s \e}^{\s} (r) \s_0 ]>
\eeq

\beq\label{see}
-\de_h C_{\s \e}^{\e}(r) <\e_0 \e_{R'}>=
\int' d^2z <\s_z \e_{R'}[ \s_r\e_0 -C_{\s \e}^{\s} (r) \s_0
 -C_{\s \e}^{\s^1} (r) \s^1_0  ]>
\eeq
(where the prime on the integrals
denotes a rotationally invariant I.R. regularization, all limits have been omitted and
we defined $\s^1\equiv L_{-1} \bar L_{-1} \s$).

It is easily realized by dimensional analysis that
 in all expressions above  no finite contribution can arise from I.R. 
counterterms: we are thus in the
 case described in Section \ref{mellin}, where computations
are easier.

To reconstruct the short distance corrections of
complete correlators we must
have the expressions of the VEV of  the lowest dimensional operators, $\s$ and $\e$.
As well known in perturbative quantum field theories and from direct Renormalization Group 
considerations, see e.g. \cite{sonoda}, composite operators VEV's evolve
by RG as
\beq\label{rg}
{d\over d l} <[\Phi_a]>= \Gamma_a^b <[\Phi_b]>
\eeq
in which $l$ is the
logarithm of the scale and  the 
matrix $\Gamma_a^b$ contains only 
positive  integer  powers of the renormalized couplings $\lambda^i$
and forbids the  mixing with higher dimensional operators (it is lower diagonal if operators are ordered with increasing dimensions).
From these properties 
and from the fact that $h$ has
 dimension $15/8$ and that it is not renormalized
it 
follows by simple 
dimensional considerations 
 that 
 $\s$ and $\e$ do not mix with other operators, 
that $\Gamma_\s^\s$, $\Gamma_\e^\e$ reduce to the fixed point 
scale dimensions and  that no logarithms appear in their VEV's. 
We can thus parameterize the VEV's as
\beq <\s>_h =A_\s |h|^{1/15}{\rm Sign}(h)\qquad <\e>_h=A_\e |h|^{8/15}. \label{a}
\eeq
(where it has been  also  kept into account the fact that $<\s >_h$ ($<\e>_h$) 
have to be odd (even) in $h$ as easily seen directly from the 
corresponding lattice expressions).

Defining ${\widehat C}\equiv C 
|r|^{{\hbox {dim}} C}$, 
$F_{A,B}\equiv {{<AB>_h}}|r|^{{\hbox {dim}} AB}$
and the scaling variable $t\equiv |h||r|^{15/8}$ we have
the following expression for the short distance (small $t$) behavior:
\bea
F_{\s \s}&=& {\widehat  {C_{\s\s}^\1 }}+A_\e {\widehat{ C_{\s\s}^\e} t^{8/15}} 
+A_\s
{\widehat{ \de_h C_{\s\s}^\s }}t^{16/15}+O(t^2) \label{ss}
\\
F_{\e \e}&=& {\widehat {   C_{\e\e}^\1} }
+A_\s {\widehat {\de_h C_{\e\e}^\s }} t^{16/15}+O(t^2)
\\
\kappa
F_{\s \e}  &=&  A_\s {\widehat { C_{\s\e}^\s }} t^{1/15}
             +{\widehat{\de_h C_{\s\e}^\1}} t+ A_\e {\widehat{\de_h C_{\s\e}^\e}} t^{23/15}+
  O(t^{31/15}) \label{se}
\eea
where $\kappa\equiv {\rm Sign}(h)$ and
\bea
{\widehat{ \de_h C_{\s\s}^\s }}&=&{1\over 32} \left({\Gamma(-1/2) \Gamma(7/4)\over \Gamma(5/4)}\right)^2 
\label{resultsss}  \\
{\widehat {\de_h C_{\e\e}^\s }}&=&0\label{zero1}\\
{\widehat{\de_h C_{\s\e}^\1}}&=& 
{1\over4} {\sin (3\pi/8) \over \sin (15\pi/8)} {\Gamma^2(11/8)\over \Gamma^2(15/8)}  
\\
{\widehat{\de_h C_{\s\e}^\e}}&=&
\sin(\pi/8) \sin (3\pi/8)
\left( 
{{15\over 8}\Gamma (-11/8)\Gamma(7/8)\over \Gamma(-1/2) }
\right)^2 \label{resultsee}
\eea
have been computed in \ref{computations} (see also \ref{conformal} for zeroth order expressions)
and  the only unknown constants, up to now are  $A_\e$ and $A_\s$.
While in general  the OPE approach leaves some {\it universal}
 unfixed constants, in this case the (approximate) value of $A_\s$ ($A_\e$)
can be obtained from the existing knowledge, see next Section.

\subsection{The missing VEV's}\label{vev}

The coefficient $A_\s$ in (\ref{a}) can be easily extracted by use of the results
of \cite{fateev} (see also references therein),
 where the integrability of the ${\cal M} (3/4)$ +
$\Phi_{12}(\equiv\s )$ was used essentially 
(in particular the Thermodynamic Bethe Ansatz and the method of
external fields have been used).

We report here Eqs. (2.9), (3.15) of \cite{fateev},
\bea
M&=&\left[{4 \pi^2\over \gamma^2 (3/16)\gamma(1/4) }\right]^{4/15}
{4 \sin (\pi/5) \Gamma (1/5) \over \Gamma(2/3)\Gamma(8/5)}h^{8/15} \non\\
&=& 4.404908579981566037 \cdots h^{8/15}
\label{mass}\eea
\bea
\epsilon&=&- \left( {M\over 2 \sin (\pi/5) }\right)^2 
{\sin (\pi /5) \over 8 \sin (\pi/3) \sin (8\pi /15)}\non\\
&=&-0.061728589822368 \cdots M^2,
\eea
where  $M$ is the minimum mass of the particle of the theory,
$\gamma(x)\equiv\Gamma(x)/\Gamma(1-x)$, 
$h>0$ in this Section  and 
the bulk energy $\epsilon$
is defined by the partition function of the theory I.R. regulated in a cylinder 
of dimensions $L,R$:
\beq
Z(h)=<e^{-\int h \s}>\sim e^{-\epsilon RL} \qquad L,R\to \infty
.\eeq

Noticing then that 
\beq
<\s>_h=<\s e^{-\int h \s}>/<e^{-\int h \s}>\sim \lim_{L,R} (-\de_h\log Z(h)/L R) 
\eeq
it follows

\beq
<\s>_h =\de_h \epsilon= A_\s h^{1/15}
\eeq
where 
\bea
A_\s&=&-{8\over 15} 
\left( {4\pi^2\over \gamma^2(3/16) \gamma(1/4)}\right)^{8/15} 
{\sin {\pi\over 5} \over \sin{\pi\over 3} \sin{8\pi\over 15}}
\left({\Gamma(1/5)\over \Gamma(2/3)\Gamma(8/15)}\right)^2\\
&=&-1.27758227605119295\cdots
\eea

On the other hand no nonperturbative informations are available
(in our knowledge)
for the VEV of operators different from the perturbation, $\s$.  

In principle, numerical estimates of the constant $A_\e$ can be obtained by comparison
with Monte Carlo simulations  on the lattice for
the
normalized  (connected) correlator 
$<\s(R) \s(0)>_c /<\s>^2$ in the  regime in which $R>>1$ 
and $t^{8/15} \propto R/\xi $ is not too big, i.e., being $\xi \propto H^{-8/15}$,
in a small field 
 ($R,\xi,H$ being lattice spacing, correlation length and  magnetic field
and the lattice spacing being assumed everywhere to be one).
In practice with the available data, \cite{rittenberg}, one has only a small window of
applicability ($\xi\sim 15.5$ for $H=.001$). To improve convergence 
and enlarge this 
window of applicability we fitted, together with the powers given in 
Eq.(\ref{ss}) also the next to leading  
terms $t^2, t^{32/15}$ 
(notice that logarithmic terms are absent 
for the same considerations given in Section \ref{isingh}).
From the known first two  coefficients in the
small $t$ expansion for the 
 continuous expression of $<\s (R) \s(0)>_c /<\s >^2$ we can fix the overall 
normalization 
(i.e. the lattice magnetization $<\s >_{{\rm lat}}$) and the correlation length $\xi$, related to continuous variables  
by $R/\xi=Mr$ ($M$ being the mass in Eq.(\ref{mass})).
Notice that the predictions (\ref{ss})-(\ref{se}) refer 
to the complete correlators,
and that the  normalization factor between the 
continuous 
spin VEV and the lattice one (physical magnetization), is negative in our conventions, see also \cite{rittenberg}.

At this point all is fixed and we get a prediction for the coefficient of 
$t^{8/15}$ ($A_{\e}$) and 
for the coefficient of $t^{16/15}$ (that actually is known
exactly). In practice we selected 
only those fit that give a good prediction for the known coefficient, fixing in this way the
above mentioned (but unknown) window
of applicability: a good representative of which is
(for H=.001) the $R=5-54$ range
that corresponds to
\beq\label{rough}
A_\e\sim 0.321
\eeq
(see Eq.(\ref{enorm}) for our normalization)
with an error of order few percent (estimated roughly from the variation
of this quantity among different "good" fits). As a byproduct we also get
\beq\label{stima}
\xi\sim 15.4 \qquad <\s>_{{\rm lat}}\sim.634\eeq
in agreement  with the estimate $\xi=.38(1)H^{-8/15}$ of
\cite{rittenberg} and $<\s>_{{\rm lat}}=1.003(2)$ of \cite{destri} (with larger lattice).

As an independent check that our estimate is correct we  can use the known
exact sum rule \cite{delfino,delfino2,delfino3}
\beq \label{sumrule}
\Delta^\s=-{2\pi h (2-2\Delta^\s) \over 4\pi <\s >} 
\int d^2 r <\s (r) \s(0)>_{h, c}
.\eeq
We  split the integral into two pieces, $Mr<\Lambda$ and $Mr>\Lambda$.
We estimated the first term by use of our expression (\ref{ss}) and 
the second  one by use of the first three terms of the
 long distance expansion 
\beq
<\s (r)\s(0)>_{h,c}\simeq \sum_{i=1}^3 {1\over \pi}{(F_i^{\s})}^2 K_0(c_i mr)
\eeq 
where coefficients $c_i$ and form factors $F_i^{\s}$ 
can be found in \cite{delfino} (and references therein).
By use of (\ref{sumrule}) we can give an expression of $A_\e$ in terms of $\Lambda$. Clearly for $\Lambda$ small we have a big error from the
 approximated long distance
 expansion, while for small $\Lambda$ we have a big error from the approximated small distance expansion. 
If (by a minimal sensitivity criterion)
 we choose the value of $\Lambda$ that minimizes $A_\e$ 
($\Lambda\sim 1.5)$ we obtain
the value $A_\e\sim .322$,
 with an error of order 
$2\%$ (estimated by repeating the procedure with only one mass term)
that is compatible with  Eq.(\ref{rough}).

\subsection{Comments}\label{comments}

A first glance at our results shows  that a coefficient of the expansion
for $<\e \e>$  
(\ref{zero1})  vanishes  nontrivially 
(unexpectedly from the point of view of the selection rules of the conformal field 
theory).
This might sound strange, but one  must bear in mind that 
the complete theory is an integrable model 
with powerful symmetries: we can thus interpret the found zero as a 
(short distance) signal of these symmetries
and possibly of the existence of nontrivial differential equations 
satisfied by
the correlator (analog of Painlev\'e equations for 
spin-spin correlators in thermal perturbations of the Ising model, \cite{mccoy}). 

Another important feature of our prediction is the presence of fractional 
powers of the coupling (or equivalently of the scaling variable 
$t=|h| |r|^{8/15}$):
in the
OPE approach these powers come out naturally from the VEV of operators
(times some integer power of $t$ coming from the Wilson coefficients, analytic in $t$).
In particular the fractional correction
 $t^{8/15}$ that is found   in (\ref{ss}) for the $<\s\s>$ 
correlator is 
expected if one  introduces as infrared cutoff in 
perturbative expansion
 for the correlators the correlation length $R_c\sim |h|^{-8/15}$ (see \cite{pol,dot2}). 
The existence  of this  fractional contribution
to the spin spin correlator can be recovered
from    the results  of \cite{zinn} (based too on OPE and Callan-Symanzik
equations in the $4-\epsilon$ framework)
by inserting  the exact values of critical
indices given by the conformal theory.   

Also it is interesting to see that the scaling of the $<\s\e>$ correlator
(vanishing at   the critical point), is
dominated by a fractional power, $t^{1/15}$, coming from the VEV $<\s>$.

We  must remark here that our results are
 in {\it disagreement} with an existing
expression of
the short distance behavior of spin spin correlator 
in the same context, \cite{dot1}     
coming from the effort  to apply in this case the 
perturbative expansion for correlators developed in \cite{dot2} and to obtain
the expected fractional powers of $h$ as well.
 Main difference in \cite{dot1} with respect to our 
(\ref{ss}) is the presence of 
powers of $\log t $ in the first terms 
  that, in our mind, are not 
motivated due to the absence of renormalization 
of lowest dimensional operators $\s, \e$ (and of $L_{-2}{\bar L}_{-2}\1$
 as well,
 that is 
the next scalar operator that  can have a nonzero VEV
 in this theory,
as shown in \cite{zamo,sonoda}), 
see discussion near Eq.(\ref{rg}). Also the  power $t^{24/15}$ 
present there is unmotivated from the OPE  point of view
(there is no corresponding VEV).

In Figure 1 we give a comparison of the numerical data 
for $<\s \s>$ with our 
prediction Eq.(\ref{ss}) (without the higher order terms added
in previous Section  to improve  the fit) 
where the estimated values
(\ref{rough}) as well as (\ref{stima}) are used.
What appears immediately is that the convergence of the expansion is  quite
 good:
with the first order corrections to Wilson
coefficients, we can achieve a relative error of order $2\% $ at $R/\xi=Mr\sim 1$. 
 
No numerical results exists in our knowledge for $<\s\e>$ and $<\e\e>$.
Nevertheless in Figure 2 we show a comparison between our (short distance)
prediction and the (long distance) form factor result
(up to the first eight terms),  \cite{delfino2}, in the
case of $<\s\e>$. It appears
from Figure 2  that there is a good agreement
of two methods  in the intermediate region
$ 1<Mr<1.5 $ and a reasonable evidence of convergence of the long distance 
expansion 
 towards our result in the region $Mr<1$, as it is
 expected. This agreement could be regarded as an independent test of
the  form factor clustering property that have been assumed in
\cite{delfino2} to get the form factors
for $<\s\e>$ and $<\e\e>$.
 (The origin of this property will be explained in \cite{delfino3}).

\section{Conclusions}\label{conclusions}

We computed  the short distance behavior of all the
correlators of primary fields of the 
critical two dimensional Ising model perturbed by a magnetic field
up to $O(t^2)$ excluded, $t= |h||r|^{15/8}$ being the scaling variable.
The I.R. finite OPE approach, extensively developed in \cite{gm} has been used.
This method does not give 
 informations about VEV's by itself.
An unfixed constant for  $<\s>$
has been obtained by use of nonperturbative results 
coming from the  integrability of the model (TBA), while  
by comparison with the existing numerical data we got  the
nontrivial estimate  (\ref{rough}) of $<\e >$ (not predicted from TBA).

We point out  once more  that with the OPE approach the expected
\cite{pol}  fractional  powers
of the coupling $h$ are {\it naturally} obtained from the operators VEV's,
while in  alternative perturbative direct estimates of the correlators 
(I.R. regularized in some way) these non integer powers are absent.
In particular we found immediately the expected  $t^{8/15}$
correction to the spin spin correlator, \cite{pol}, and we predicted that the 
dominant term of $<\s \e>$ is the fractionary term $t^{1/15}$ (that should
 be observable  in adequate lattice simulations).

The result for $<\s \s>$ when compared with the available data 
signal a good convergence of the approach: this encourages 
 to search some
resummation technique that
could enlarge furthermore  the convergence radius.
Moreover the use of exact  sum rules to extract informations from both short and long distance approximations that has been done in Section \ref{vev},
should stimulate the research in this direction, with the 
final goal of connecting the two regimes.

We conclude by  emphasizing  that the OPE approach, being
very general (integrability of the theory is not necessary to compute
Wilson coefficients)
and fastly convergent, can always be seen as a bridge towards 
the study of non integrable perturbations
and as a test for the ansatz behind  integrable theories
predictions  (as done in Section \ref{comments}). 
 It is furthermore  an open question if the powerful Coulomb gas technique
of
\cite{dot3,dot4} and the  IR regularization of
\cite{dot2} 
could  be married with the OPE approach to reach higher orders of the
 expansion
in this and other statistical models.

\smallskip {\bf Acknowledgements:}
One of the author (R.G.) is indebted with D. Bernard
and J. Zinn-Justin for useful discussions and
careful reading of the manuscript.
 The authors acknowledge  J. Bros, M. Caselle, G. Delfino, P. Di Francesco,
 G. Mussardo, S. Nonnemacher, Al. B. Zamolodchikov,  J.B. Zuber
for interesting informations and discussions.
Authors also thank   G. Delfino and P. Simonetti for sending them a table of the (eight terms) form factors prediction for the spin energy correlator. 
The work of R.G. is supported by a TMR EC grant, contract N$^o$ ERB-FMBI-CT-95.0130.
R.G. also thanks  the INFN group of Genova for the kind hospitality.
\appendix

\section{Computation of the integrals.}
\label{computations}
First of all we give the general expression for the Mellin transform
(see Eq.(\ref{mellint}))
 with respect to $m$ (I.R. cutoff) of the 
integral:
\beq\label{integral}
I_{\alpha,\beta,\gamma}(m;x)
=\int d^2w \Theta(m|w|) |w|^{2\alpha} |w-x|^{2\gamma} |w-1|^{2\beta}
\eeq
in which $m>0$, $x\in (0,1)$, $\alpha,\beta,\gamma >-1$ (to have local integrability) and we will fix the I.R. cutoff function $\Theta(t)=e^{-t}$  
(such that its Mellin transform is ${\tilde \Theta}(s)=\Gamma(s)$).
By use of the convolution theorem (\ref{mellinc})
it easy to obtain 
\beq
{\tilde I}_{\alpha,\beta,\gamma}(s;x)= 
\Gamma(s) D(\alpha-s/2,\beta,\gamma, x) \label{bella}
\eeq
where the integral
\beq
D(a,b,c,x)\equiv \int d^2w  |w|^{2a} |w-x|^{2c} |w-1|^{2b}
\eeq
has been computed in \cite{dot4} by use of  contour deformation  
in the variable ${\rm Im} w$ and found to give:
\beq
D(a,b,c,x)={S(a) S(c) \over S(a+c)} |I_{0x}|^2 +{S(b) S(a+b+c) 
\over S(a+c)} |I_{1\infty}|^2
\eeq
where
\bea
S(a)&\equiv& \sin \pi a \non\\
I_{0x}&\equiv&\int_0^x du u^a(x-u)^c (1-u)^b\non\\
&=&x^{1+a+c} {\Gamma(a+1)\Gamma(c+1) \over \Gamma(a+c+2)} F(-b,a+1;a+c+2;x)
\non\\
I_{1\infty}&\equiv&\int_1^\infty du u^a(u-x)^c (u-1)^b
\non\\
&=&{\Gamma(-a-b-c-1)\Gamma(b+1) \over \Gamma(-a-c)} 
F(-a-b-c-1,-c;-a-c;x) \non
\eea
and $F\equiv {}_2F_1$ is the Hypergeometric function (see e.g. \cite{pbm}).
From the Mellin transform (\ref{bella}) we can obtain the asymptotic expansion 
of the original integral (\ref{integral}) when 
$m\to 0$, by use of (\ref{asint}).

All the required integrals in (\ref{sss}-\ref{see}) can be then computed by 
use of  the general expression (\ref{bella}). See \ref{conformal} for 
the conformal correlators.
In our particular case we know that the I.R. counterterms cannot give $m^0$ corrections, see Section \ref{isingh}
 so that the derivative of the Wilson coefficient is obtained by 
extracting the $m^0$ contribution  of the naive perturbative term 
(the first of right hand side) in 
(\ref{sss}-\ref{see}), i.e. by computing the residue of its 
Mellin transform in $s=0$. Notice that due to the absence of $m^0\log m$ 
I.R. counterterms, there cannot be $\log m$ factors in the naive term
 and correspondently  its Mellin transform 
 will have only simple poles. 
While (\ref{se1}-\ref{see}) are trivially reconduced to (\ref{bella}),
 some few words must be spent for (\ref{sss}), i.e. for the Mellin
transform of four spin correlator (\ref{4s}). 
In the limit $R'm \to \infty$ ($R'$ being the argument of the boundary operator
$\s_{R'}$)  
we have in the correlator 
the simplification $x={R'-z\over (R'-1) z}={1\over z}$ ($r=1$)
so that it is  better to change variable and obtain (in notation of 
the convolution theorem (\ref{mellint})):
\beq\label{xint}
{\tilde G} (1-s)=\int d^2x |x|^{s-4} |1-x |^{-2\delta }
(|{1+\sqrt{1-x}\over 2} |+|{1-\sqrt{1-x}\over 2} |)
.\eeq 
As explained in Section \ref{mellin}
we have introduced an additional parameter $\delta$ to justify 
the use of convolution theorem. The analytic continuation
to $\delta=1/8$ 
 will give the wanted Mellin transform 
of the four spin correlator with exponential cutoff.
The final step to reduce (\ref{xint}) to (\ref{bella}) is the simple
substitution $w={\sqrt {1-x}}$ and the observation that the integrand
is invariant for  $w\to -w$.
By use of these considerations Eqs.(\ref{resultsss}-\ref{resultsee})
follow.

As a general check of
the regularization independence our approach,
 we report the alternative 
derivation of
$\de_h C_{\s \s}^\s$ obtained regularizing the integrals by 
restricting them to $|z|<R$. We will not use the Mellin transform technique
and we will keep explicitly the  I.R. counterterms to show
explicitly how cancellations
work.

 In the limit $R'/R \to \infty$ (being $x=1/z$) 
we  can rewrite 
 (\ref{sss})  as
\beq
 -\widehat{\de_h C_{\s \s}^{\s}}
 =\int_{|x| > 1/R} {d^2x \over |x|^4}
[ {|\gp(x)|^2+ |\gm(x)|^2 } -1 -{1\over 4} |x| ],
\label{sssexpression}\eeq
where we omitted 
 the remaining overall limit $R \to \infty$ in left hand side and defined
\beq
g^{\pm}(x)\equiv \sqrt{{1\pm\sqrt{1-x} \over 2 (1-x)^{1/4}}}
.\eeq

Unfortunately we have not a closed expression for the integral
(\ref{sssexpression}). 
However the calculation can be performed by series,
 splitting the integral 
in two regions, ${1\over R}<|x|<1$ and $|x|>1$. 
The contribution of the 
first region
is then obtained expanding 
\beq
\gp(x)=\sum_{j=0}^{\infty} \gp_jx^j 
\qquad \gm(x)=\sum_{j=0}^{\infty} \gm_j x^{j+1/2} 
\eeq
(note also that  $ (g_{\pm} (x))^*= g_{\pm} (x^*)$),
 exchanging the series with  the integral and performing simple integrals of
powers of $x$ $x^*$. At this stage the role of I.R. counterterms is 
essential to give a finite $R\to \infty $ limit.
The contribution of the second region is obtained
after making the change  of variable $x=-{1\over w^2}$, 
 from  which we obtain
\beq
2\int _{|w|<1}{d^2 w} |w|^{3/2} |h(w)|^2
\eeq
\beq
h(w)\equiv \sqrt{{w+\sqrt{1+w^2}\over (1+w^2)^{1/4}}}
\eeq
for the sum of $g^{\pm}$ contributions (integrals of I.R. counterterms are
easily performed).
Again developing 
\beq
h(w) =\sum_{j=0}^{\infty} h_j w^{j} 
\eeq
the integral can be easily performed by series.

The final result is
\beq
 \widehat{\de_h C_{\s \s}^{\s}} =-2\pi \left({1\over 64}+
\sum_{j\ge 2}({|\gp_j|^2\over 2j-2} +{|\gm_j|^2\over 2j-1})
-{3\over 4} +{4\over 7} +{1\over 11}+\sum_{j\ge 2} {2 |h_j|^2\over 2j+7/2}
\right)
,\eeq
where the coefficients $g^{\pm}_j, h_j$ satisfies the recursive relations
\bea
 & &(j+1) (j+\half)\gp_{j+1}
-2 j (j-{1\over 8}) \gp_j+
(j^2-{7\over 4}j+{45\over 64})\gp_{j-1} =0\non\\
 & &(j+1) (j+{3\over 2})\gm_{j+1}
-2 (j+\half) (j+{3\over 8}) \gm_j+
(j^2-{3\over 4}j+{5\over 64})\gm_{j-1} =0\non\\
& &
(j+1)(j+2)h_{j+2}+j(2 j-\half) h_j +( j^2-{7\over 2}j +{45\over 16}) h_{j-2}=0
\label{recursive}\eea
due to the existence of the following differential equations 
satisfied by $g^\pm , h$:
\bea
& &(x^3-2 x^2 +x) {g^{\pm}}^{''}
 +({5\over 4} x^2 -{7\over 4} x+\half ){g^{\pm}}^{'}
-{3\over 64} x {g^{\pm}}=0 \non\\
& &(1+2w^2+w^4) {h}^{''} +{3\over 2}(w^3+w){h}^{'}-{3\over 16} w^2 h=0
.\eea

As they stand the series are very slowly convergent,
($g^{\pm}_j , h_j \sim j^{-7/8}$) so to increase the convergence of the series
we added and subtract to the $j^{th}$ term its asymptotic powerlike 
behavior\footnote{The first two terms are  obtained from the contribution of the 
singularities nearest to zero (Darboux theorem, see \cite{darboux,wong})
while the others can be conveniently extracted by expanding in 
$j^{-7/8+k/2}$ 
the recursive relations (\ref{recursive}) up to the desired order $k$.}
that can be resummed and gives 
well known Lerch functions 
($\Phi(z,s,a)=\sum_{n+a\neq 0} {z^n\over (n+a)^s}$, $n=0,1,\cdots$) that are obtained with high
precision from Mathematica.

After numerical resummation 
 we obtain then
\beq \label{sssfinal}
\widehat{\de_h C_{\s \s}^{\s}} =.403745631213448123\cdots
\eeq
that agrees up to the obtained precision with the analytic result
(\ref{resultsss}).

\section{Conformal correlators}\label{conformal}
To fix notations we
 give here the expressions for the fixed point (conformal theory)
quantities  involved in our computations. See \cite{zuber,mz} for their derivation.

Wilson Coefficients:
\beq\label{enorm}
{C_{\e\e}^1(r)}={1\over 4 \pi^2 |r|^2} 
\eeq
\beq 
{C_{\s\s}^1(r)}={1\over |r|^{1/4}} \qquad
{C_{\s\s}^\e (r)}={\pi |r|^{3/4}} 
\eeq
\beq
{C_{\s\e}^\s(r)}={1\over 4\pi|r|} 
\eeq

Useful correlators:
\beq
<\s(z_1) \s(z_2) \e(z_3)>
={1\over 4\pi} {|z_{12}|^{3/4}\over |z_{13}||z_{23}|}
\eeq

\beq
<\s(z_1) \s(z_2) \e(z_3) \e(z_4)>=
{|z_{12} (z_{32}+z_{42}) -2 z_{32} z_{42}|^2    \over
16 \pi^2 |z_{42} z_{32} z_{41} z_{31}| |z_{43}|^2 |z_{12}|^{1/4}}
\eeq

\beq\label{4s}
<\s(z_1) \s(z_2) \s(z_3) \s(z_4)>=
|(1-x )z_{12} z_{34}|^{-1/4}
(|{1+\sqrt{1-x}\over 2} |+|{1-\sqrt{1-x}\over 2} |)
\eeq
where $x=z_{12} z_{34} / (z_{13} z_{24})$, $z_{ij}=z_i-z_j$.

\newpage
{\bf Figure Caption:}

\smallskip
{\it Figure 1} Values of $<\s\s>$ from lattice simulation (dots) and
as predicted from OPE approach (solid line)
versus R (lattice position)
for $H=.001$ ($\xi\sim15.5$).  

\smallskip
{\it Figure 2} Estimates of ${<\s(r)\e(0)>\over <\s> <\e>}$ versus $Mr$:
comparison between the OPE approach (solid line) and the form factor method
with one (dashed line) three (dot-dashed line) and eight (dots) terms. 
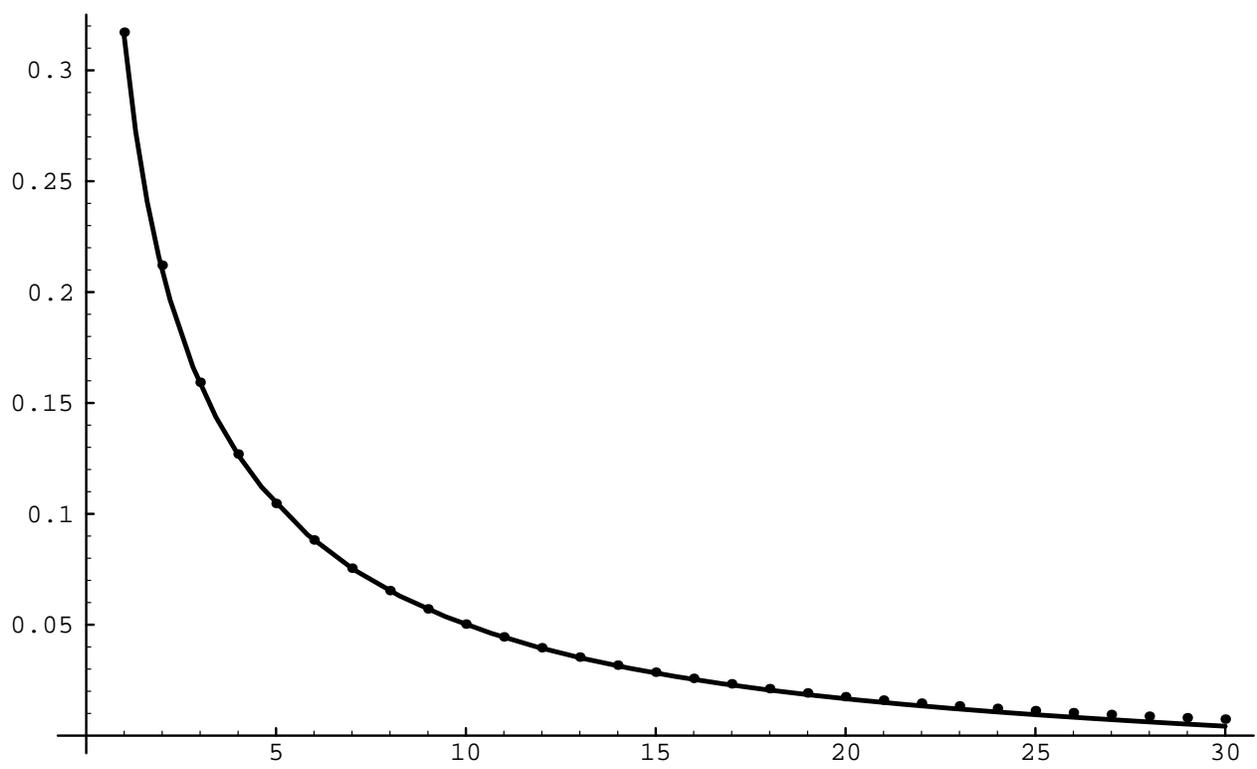
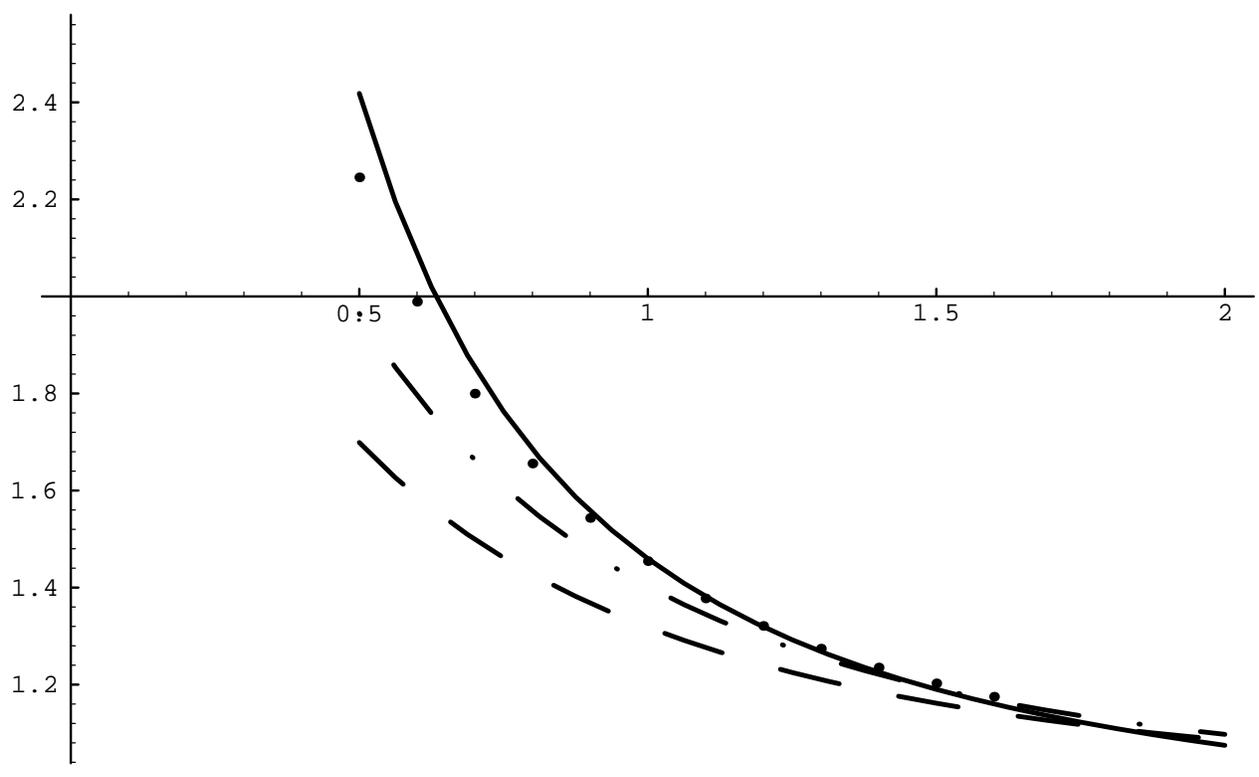
\end{document}